\documentstyle[12pt,aaspp4]{article}

\def\mathnew{\mathsurround=0pt}
\def\simov#1#2{\lower .5pt\vbox{\baselineskip0pt \lineskip-.5pt
        \ialign{$\mathnew#1\hfil##\hfil$\crcr#2\crcr\sim\crcr}}}
\def\simgreat{\mathrel{\mathpalette\simov >}}

\def\simless{\mathrel{\mathpalette\simov <}}

\def\g{{$\gamma$}}

\begin{document}
\title{ Constraints on the VHE Emissivity of the Universe
from the Diffuse GeV Gamma-Ray Background }
\author{Paolo S. Coppi}
\affil{Department of Astronomy, Yale University, P.O.
Box 208101, New Haven, CT 06520-8101}
\authoremail{coppi@astro.yale.edu}
\author{Felix A. Aharonian}
\affil{ Max-Planck-Institut f\"ur Kernphysik, Postfach 103980, D-69029 
Heidelberg, Germany}
\authoremail{aharon@fel.mpi-hd.mpg.de}
\centerline{(Shortened version submitted to {\it Ap.J. Lett.})}

\begin{abstract} VHE (Very High Energy, $E \gtrsim$ 100 GeV) radiation
emitted at cosmological distances will pair produce on
low-energy diffuse extragalactic background radiation 
before ever reaching us.
This prevents us from directly seeing most of the VHE emission 
in the Universe.  However, a VHE \g-ray that pair produces 
initiates an electromagnetic pair cascade. At low energies,
this secondary cascade radiation 
has a spectrum insensitive to the spectrum of 
the primary $\gamma-$radiation and, unlike the original VHE radiation,
{\it is observable}.
Motivated by new
measurements of the extragalactic MeV-GeV diffuse \g-ray background,
we discuss the constraints placed on cosmological VHE source
populations 
by requiring that the cascade background they produce
not exceed the
observed levels. We use a new, accurate
cascading code and pay particular attention to the dependence
of the constraints on the diffuse cosmic background 
at  infrared/optical wavelengths. Despite considerable uncertainty
in this background,
we find that robust constraints may still be placed on the integrated
emissivity of potential VHE sources in the Universe. The limits 
are tighter than those obtained by considering cascading 
on the microwave background alone and restrict
significantly, for example, the parameter space available for the
exotic particle physics scenarios recently proposed to explain the
highest energy cosmic ray events. If direct emission from blazar 
AGN in fact accounts for most of the observed GeV background, 
these limits strengthen and rule out AGN emission scenarios 
which produce significant power above $\sim$ 300 GeV.
\end{abstract}

\section{Introduction}

Photon-photon pair production of high energy
\g-rays on the 
diffuse extragalactic background radiation (DEBRA)
significantly limits
the distance that such \g-rays can propagate (\cite{Nikishov},
\cite{GouldS}).
For \g-rays energies above a few TeV, this distance is almost
certainly $\simless 100$ Mpc (\cite{steck1}). Thus, most
of the VHE Universe is  not visible to us.
Nevertheless, we can detect the presence of VHE
emission via the lower-energy, 
secondary radiation produced in the cascades initiated by the VHE
photons.
(The energy carried by absorbed VHE \g-rays is
not lost; the electron-positron pairs produced 
create new \g-rays by
inverse Compton scattering on  background field photons
and can trigger an electromagnetic pair cascade.) In principle,
this radiation could be detected from an individual
source as a halo about the source (\cite{AhC}, \cite{PM}), 
but an easier way
to detect or constrain VHE emission 
is to look for the diffuse cascade background produced by
the ensemble of all VHE sources in the Universe. In particular, 
the spectrum
of this secondary cascade radiation is rather insensitive to the spectrum
of the primary VHE radiation, and therefore the total level
of the cascade background acts as a particle detector calorimeter,
allowing us to measure the total VHE energy input into the Universe.
This was first pointed out in the context of detecting
an extragalactic population of high-energy cosmic rays
(\cite{Wdow}).
%% \cite{Strong}, and later \cite{Halzen}).
However,
cosmic rays 
are not the only possible sources of VHE radiation.
The physical conditions in powerful, non-thermal extragalactic
sources like Active Galactic Nuclei (AGN) and Radio Galaxies
could lead to the acceleration of particles to very high
energies (e.g., see the review of Hillas 1984), 
and depending on exactly where acceleration occurs, AGN 
can easily contain enough target matter (e.g., strong
photon fields) to convert efficiently this particle energy into high
energy radiation. Indeed, the nearby blazar AGN
Mrk 421 and 501 do emit TeV \g-rays. 
Also, several ``exotic'' particle physics/early Universe scenarios
have been proposed that lead to the
generation of extremely high energy 
charged particles and photons at late times
(e.g., from decaying primordial black
holes or GUT-scale particles and topological defects) .
% ; see \cite{Geddes} for a recent example). 
These scenarios can
produce a large VHE cascade background, and
tight constraints on them result from requiring that: (i) the level of the 
cascade background today not exceed the observed extragalactic 
\g-ray background, and (ii) the level of the background at 
high redshifts not be enough to alter primordial nucleosynthesis (e.g., 
see the review by \cite{Ellis2}). 

In this paper, we re-examine and quantify the constraints 
placed on a cosmological VHE source population by requiring
that its cascade background not exceed the \g-ray background observed
today. We are motivated by two considerations. First, EGRET has
provided us with a determination of the diffuse
extragalactic background at $1-10$ GeV
(\cite{Fichtel}). This is the energy range 
where the expected cascade backgrounds are typically most insensitive to 
model parameters and from which we thus obtain the most robust
constraints. Second, the prior work we are aware of 
either does not accurately
compute the cascade radiation spectrum or only considers cascading
on the microwave portion of the DEBRA. The use of a realistic DEBRA, 
i.e., one that includes an infrared/optical (IR/O) component,
is key as it leads to much tighter cascading  constraints.
Except for the microwave background (MBR),
our current knowledge of the DEBRA and its 
evolution in time is limited.
In \S 2 of the paper, we illustrate the uncertainties in the
propagation of VHE photons corresponding to these uncertainties in 
the DEBRA. These can be significant and should not be forgotten
when evaluating claims based on calculations of
VHE photon absorption and cascading.
In \S 3, we calculate the cascade backgrounds expected for 
different IR/O background fields and VHE source population models.
Using decaying particles from topological defects (hypothesized
as a possible source for the highest energy
cosmic rays) and AGN as example populations,
we show how interesting and straightforward astrophysical source
constraints can be derived -- {\it despite} the uncertainties in the
DEBRA. We 
summarize our results in \S 4 and discuss their implications
for future high energy \g-ray experiments.

\section{VHE Photon Propagation Through the Cosmic Background
Radiation}
The distance a VHE photon propagates in the Universe
is determined by the intensity of the intervening DEBRA.
For small emission redshifts,
the photon propagation length at energy
$E$ is $\lambda(E) \approx 2.5 \epsilon_s 
[\epsilon_s U_\epsilon(\epsilon_s)]^{-1}$ where $\epsilon_s(E)
\approx 0.25 (E/1\,{\rm TeV})^{-1}$ {\rm eV}
and $\epsilon_s U_\epsilon(\epsilon_s)$
is the background energy density at energy $\epsilon_s(E)$ 
(e.g., \cite{Hert}). In other words,
there is a rough one-to-one mapping between $\lambda(E)$ and the 
background intensity at $\epsilon_s(E).$  A determination of
$\lambda(E)$ via the detection of an absorption cutoff in a
spectrum thus measures the background at $\epsilon_s(E)$
(and {\it only} $\epsilon_s(E)$!) --
a possibility that has aroused much interest given the 
observational difficulties in extracting the extragalactic IR/O
background from the galactic/solar system foregrounds. Conversely,
to understand VHE photon propagation, we require accurate knowledge
of the DEBRA, in particular at IR/O energies.
Unfortunately, direct measurements of the IR/O 
background (\cite {Puget}) are at best preliminary.
Theoretical estimates for the DEBRA exist (e.g., 
\cite{Franceschini}) but are also rather uncertain. Consequently,
our estimates for the propagation lengths
of VHE photons emitted today are similarly uncertain.
We summarize these uncertainties in Fig. 1. 
They are not small.
In the ultra-high photon
energy range
($\epsilon \simgreat 10^{20}$ eV), they arise from problems in
determining the low-frequency cutoffs of extragalactic radio sources.
%(e.g., see the recent paper of \cite{ProtB}, but note that their
%calculated background appears to exceed the total observed background
%below 1 MHz). 
In the GeV-TeV range, the uncertainties reflect our
poor understanding of galaxy formation and evolution (
e.g., see \cite{MnP}, \cite{Madaup} for detailed discussions).

For a VHE photon emitted at redshifts $z_{\rm emit} \gtrsim 0.1,$
an added complication arises. As it propagates, the photon's 
energy is redshifted and at any given redshift $z^\prime,$ 
the photon interacts most strongly with 
local background photons of  energy
$\epsilon_s \propto {(1+z^\prime) \over (1+z_{\rm emit})}E_{\rm emit}^{-1}$
where $E_{\rm emit}$ is the energy of the 
VHE photon at $z_{\rm emit}.$
Therefore,
to estimate the probability
for this photon to be absorbed,
we need good estimates for the DEBRA over a range of energies 
{\it and} over a range of redshifts. A common assumption
is that the  DEBRA was produced in a burst
at $z_{\rm burst} =\infty$ like the MBR,
so that the DEBRA photon density also scales
as $n(\epsilon,z)=(1+z)^3n[\epsilon/(1+z),0]$ (e.g.,
\cite{steck1}, \cite{Biller}). However, 
galaxy emission, the likely source
of the  IR/O DEBRA, evolves in time as galaxy stellar populations
evolve, and galaxies still emit significant amounts of light today
(i.e., their emission burst is not over).
The exact epoch of galaxy formation is highly controversial,
but formation redshifts as low as $z_{\rm form} \sim 1-3$ are
commonly considered. In other words, the evolution of the non-MBR DEBRA
could deviate considerably from the $(1+z)^3$ law. 
Consequently,  uncertainties in the DEBRA actually have a greater
impact on \g-ray propagation than implied by Fig. 1. 
This is not always appreciated. In Fig. 2,
we show for several DEBRA evolutionary scenarios the observed cutoff energy, 
$E_{\rm cut},$ in the spectrum of a VHE source located at redshift $z.$ In
curves (i) and (vi) of Fig.2, we show $E_{\rm cut}$ for roughly the minimum
(Tyson 1995) and maximum (Dwek \& Slavin 1994) allowed IR/O levels today:
$\epsilon^2 n(\epsilon,0) = 1\times 10^{-3}\  {\rm eV}\, {\rm cm}^{-3}$ for
(i) and $\epsilon^2 n(\epsilon,0)=1 \times 10^{-2}\ {\rm eV}\,
{\rm cm}^{-3}$ for (vi), where for an order of magnitude
estimate, we assume the background goes as $n(\epsilon)
\propto\epsilon^{-2}.$ In both cases, we 
scale the DEBRA back in redshift as $n(\epsilon,z) = (1+z)^3 
n[\epsilon/(1+z),0],$  and we assume $n(\epsilon,0)$ has {\it no} 
optical/UV cutoff (e.g., a Lyman limit).  While convenient, note
that obtaining the DEBRA at high redshifts in this way,
i.e., extrapolating it from the DEBRA today, 
can be dangerous. Most likely, the today's DEBRA
is a complicated superposition of light emitted
at various redshifts. By extrapolating from the current DEBRA, we
make an implicit, but often unrealistic and unphysical
assumption about the spectra of the sources contributing to the DEBRA,
especially when the source spectra contain sharp cutoffs, e.g.,
a Lyman limit. As an example, assume that because of the 
Lyman limit, the current DEBRA cuts off sharply above, say, 5 eV.
Extrapolating to $z=2,$ we now find a DEBRA that cuts
off sharply only above  15 eV, and we could conclude that the DEBRA
at that epoch was due to sources with strong emission above 
the Lyman limit (13.6 eV) --  rather unlikely if the sources are
galaxies. Now, the
absorption cutoff in distant sources like AGN depends 
critically on the DEBRA intensity at optical/UV  energies 
(i.e., near the Lyman limit), so one must be very
careful about the optical/UV DEBRA evolution.
In curve (v), we compute
$E_{\rm cut}$ for the DEBRA of curve (vi) except that we
assume the DEBRA was produced in a burst at $z_{\rm burst}=5$
and that no photons were emitted 
above the Lyman limit during the burst.
In this case, the DEBRA at $z<z_{\rm burst}$
still scales as $(1+z)^3$, but there are no photons present
above $13.6(1+z)/(1+z_{\rm burst})$ eV.
Note the large discrepancy between curves (v) and (vi) at high $z.$

Curves (i) and (vi) are probably fairly good as absolute upper and lower 
bounds for $E_{\rm cut}$ at a given $z,$  but the assumptions underlying
them are not very realistic. To give better examples of 
the effects of DEBRA evolution, we assume that the IR/O DEBRA is 
the integrated light from galaxies, and that galaxies
have two distinct spectral components: direct/optical UV emission
from stars, and IR emission from dust which reprocesses the starlight.
We allow the component luminosities to vary (independently)
with redshift, but we keep their spectral shapes fixed and choose
them to match the intermediate age spectra in Mazzei, Xu, \& DeZotti (1992).
This prescription can reproduce fairly well more sophisticated 
calculations and allows one to efficiently explore a variety of galaxy
evolutionary scenarios. In curve (iv), we show the results for a 
scenario of the type proposed  in 
Franceschini {\it et al.} (1994) where galaxies and stars form very early 
in the universe, and the DEBRA is dominated by light produced during 
the initial burst of star formation. The calculation shown assumes 
most stars/galaxies were formed at a formation redshift 
$z_{\rm form} \sim 5,$ and the 
intensity and evolution of the spectral components were adjusted 
to give roughly the IR/O DEBRA spectrum (for $z=0$) shown in 
Franceschini {\it et al.} 1994. In curves (ii) and (iii), we show
what might happen if galaxies are instead formed at later times, as
suggested by cosmological numerical simulations and as
discussed by  MacMinn \& Primack (1995). In curve (ii), we
assume galaxies form late, at $z_{\rm form} \simless 1,$ and 
adjust the spectral
component intensities and intensity evolution to roughly match those
in the HCDM (Hot-Cold Dark Matter)-based calculation of MacMinn \& Primack
(1995). In curve (iii), we assume galaxies form at intermediate
redshifts $1 \simless z_{\rm form} \simless 3$ and 
adjust the component intensity evolution
to match the CDM (Cold Dark Matter)-based calculation of 
MacMinn \& Primack (1995).  Although
the models shown probably do not produce enough UV light
(e.g., as compared to  Madau \& Phinney 1996) 
and $E_{\rm cut}$ could be somewhat lower at high $z,$ the
curves in Fig. 2 should be indicative of the large
range of possibilities due to the current uncertainties
in the DEBRA. 
Consequently, one should be wary of extrapolating a determination
of $E_{\rm cut}$ at one redshift to other redshifts without extra information
(the curves in Fig. 2 intersect!). For example, 
an exact measurement of the IR/O DEBRA today would not be
sufficient to allow a precise determation of the Hubble
constant via the absorption cutoffs in distant ($z > 0.1$) sources.
Finally, one should be careful when claiming an unidentified source, 
e.g., a \g-ray burst, is closer than a certain redshift because it 
shows no VHE absorption.

\section {The Cascade Background from a Population of VHE Sources}

While the IR/O DEBRA uncertainty is significant, it is not
enough to allow VHE photon emitted  above $\sim$ 1 TeV to propagate
more than a few hundred Mpc (see Fig. 1; and note that
the DEBRA intensity generally increases with redshift). Similarly,
cascading typically takes less than 
a few hundred Mpc to reprocess an absorbed VHE photon's
energy into that of many photons with 
individual energies well below a TeV. Thus, for cosmological
source populations like AGN that span Gpc distance scales,
it is  a reasonable approximation to assume a
VHE photon is transformed instantaneously into sub-TeV photons.
This is the primary reason VHE cascading is such a powerful diagnostic:
essentially {\it any} energy emitted above 1 TeV re-emerges below 1 TeV, 
where we can detect it. The second is that for  VHE photon
energies above $\sim 1$ TeV, the cascade spectrum at lower energies
is very insensitive to the initial VHE photon energy distribution 
and only weakly
sensitive to the details of the IR/O background distribution (which 
determines exactly where below 1 TeV the energy ends up, i.e., $E_{\rm cut}$).
The latter point, recognized as early as Strong, Wdowczyk, \& Wolfendale
(1973), is key but has not been widely appreciated. 
The spectrum for a cascade started at $z_{\rm emit}$ goes roughly as
$dN/dE \propto E^{-1.5}$ for $E<E_{b}$ and as $E^{-\alpha_\gamma}$
for $E_{b} <E<E_{\rm cut}(z_{\rm emit})$ where $E_{b}
\sim [E_{\rm cut}(z_{\rm emit})/1{\rm TeV}]^2$ GeV 
and typically $\alpha_\gamma \sim 1.8-2$ (Coppi \& K\"onigl 1996)
The portion of this spectrum which is relevant to 100 MeV-GeV observations
and also contains most of the cascade energy is the $\sim E^{-2}$ component,
and not as usually believed, the $E^{-1.5}$ component.
Changing the IR/O background changes $E_{\rm cut}$ but not the total
observed cascade energy. Hence, the cascade photon spectrum
amplitude goes (roughly) as $\sim \ln(E_{\rm cut}/E_{\rm b}),$ i.e., 
it is rather insensitive to $E_{\rm cut}$ and thus the IR/O background.
Now, the spectrum from a population of sources
is simply the sum of cascade spectra for a range of $z_{\rm emit}.$ Below
$\bar E_{\rm cut},$ the cutoff energy for the average $z$ of 
the VHE source population, the summed cascade spectrum then
goes roughly as $\sim E^{-2}$ and is similarly insensitive 
to the IR/O background.
(However, the spectrum above $\bar E_{\rm cut}\ ${\it does} depend
strongly on the background.) If we can measure or constrain
the \g-ray background below $\bar E_{\rm cut},$ we can robustly
constrain the VHE source population luminosity.

The preceding arguments need to be backed
up by a rigorous calculation. We have developed a new code that 
solves the exact cascade kinetic equations implicitly
(see \cite{CandK}). The main advantages of our code over past
ones are: (i) it does not have to confront problems of Monte Carlo 
particle statistics and is typically much faster, and (ii)
because it is implicit, it can efficiently follow the cascade
into the Thomson regime where cascade electrons lose energy 
in very small steps and thus accurately compute the low-energy
($\simless 1$ GeV) cascade spectrum. The accuracy in this energy range
-- the range needed to compare with \g-ray observations --
has been a problem in past work (e.g., \cite{Chi1}), especially given
that small errors at high energies extrapolate to 
large errors at low energies. The cosmological (redshift) terms in the
kinetic equations are handled in the same manner as Protheroe
\& Stanev (1993).

We present two sample numerical calculations which
demonstrate the power of the VHE cascade constraint. The first, shown
in Fig. 3, is motivated by the possibility that the highest energy
cosmic rays ($E \simgreat 10^{20}$ eV) might be explained as 
decay products of massive ($\sim 10^{15}$ GeV) primordial particles/topological
defects (e.g., see \cite{Sigrev} for a recent review). 
For this calculation, 
we assume
the intergalactic magnetic field (IGMF) exceeds $\sim 10^{-11}$ G, so that
the initial cascading is suppressed by 
synchrotron losses, and we can approximate
the initial decay products as photons
of energy $\sim 10^{15-16}$ eV (e.g., see \cite{Ahabats}).
Taking the integral cosmic ray flux above $3\times 10^{20}$ eV to be 
$\approx 4\times 10^{-21} {\rm cm}^{-2} {\rm s}^{-1} {\rm sr}^{-1}$
(\cite{Fly1},\cite{Agasa}), one can easily show that the current energy 
release rate required to explain such a flux 
%(which gives us the effective $10^{15-16}$ eV photon production rate)
is $\dot Q_{\rm e-m} \approx
1.5\times 10^{-22} [(\pi^0/p)/10][E_{\rm max}/10^{24}{\rm eV}]^{1/2}\ 
{\rm eV}\ {\rm cm}^{-3} {\rm s}^{-1}$ if the observed cosmic rays
are protons, and 
$\dot Q_{\rm e-m} \approx 6.5 \times 10^{-23} [E_{\rm max}/10^{24} 
{\rm eV}]^{1/2} [\lambda_\gamma/10{\rm Mpc}]^{-1}$ if they
are photons. Here $(\pi^0/p)$ is the number ratio of 
$\pi^0$ particles to protons produced in a decay, $\lambda_\gamma$ is 
the photon absorption mean free path at $E=3\times10^{20}$ eV 
(probably between $3-20$ Mpc, see Fig. 1),
and we assume the decay product energy spectrum is 
$dN/dE \propto E^{-1.5}$ extending to energy $E_{\rm max}.$ 
(For most models, $E_{\rm max}$ lies in the range 
$10^{23}-10^{25}$ eV.)
The dependence of the cascade background on the 
IR/O DEBRA increases with stronger
decay rate evolution, but below $\bar E_{\rm cut}$ is remarkably small,
between 1-10 GeV. For the optimal case of constant comoving decay rate,
the EGRET measurement constrains the local energy release
rate to be $\simless 3 \times 10^{-23}{\rm eV}{\rm cm}^{-3}{\rm s}^{-1}.$
Thus, if the cosmic ray events above $10^{20}$ eV are caused by
protons, we can firmly rule out a primordial particle/defect
scenario as their origin.
This conclusion does not depend strongly on
the details of the IR/O DEBRA or the IGMF strength. 
If the high energy events are instead
due to \g-rays, we also find strong constraints, but with some
caveats concerning the IGMF. If the IGMF exceeds $\sim 10^{-11}$ G, 
then Fig. 3
shows the standard scenarios are at best marginally allowed. 
(This agrees with the conclusions of
Chi {\it et al} 1992, even though they appear to have significantly
overestimated the expected cascade background.) However, if the 
IGMF is weak ($\simless 10^{-11}$ G),
the initial cascading 
effectively increases the mean free path of $3\times 10^{20}$ eV photons to 
$\lambda_\gamma \sim 100$ Mpc and in this case
a decay scenario is {\it not} ruled out 
(see \cite{Sangjin}; \cite{sjlee2}).

Blazar AGN are known to emit at GeV energies and, in fact, may
dominate the observed  GeV background (e.g., 
see \cite{Steckback} and references therein). If this were the case, 
the VHE source luminosity constraints would of course tighten
considerably. In fact, they appear rather interesting when applied
to radio-loud AGN, the parent population of blazars, and to
AGN in general. As an illustration,  we repeated
(see Fig. 4) the background calculation of
Stecker \& Salamon (1996), making exactly
the same assumptions about the blazar luminosity
function. However, we also took into account
VHE photon absorption and cascading and considered various
values for the typical maximum blazar emission energy. The level
of the expected cascade background depends on the value
of the IGMF and the exact relation between the apparent luminosity
\g-ray of blazars measured by EGRET (which is enhanced by relativistic
beaming) and the intrinsic (unbeamed) \g-ray luminosity of blazars, and their 
parent population, radio-loud quasars. When the IGMF exceeds
$\sim 10^{-15}$ G, the pairs in a VHE cascade are
deflected sufficiently that 
the VHE luminosity of a blazar is effectively ``debeamed.'' 
The contribution from an individual blazar to the 
cascade background is then dramatically 
reduced, by a currently unknown factor between $\delta^2$ 
(if the typical EGRET blazar luminosity corresponds to quasi-steady
emission) and 
$\delta^4$ (if it corresponds to strong, flaring emission).
Here, $\delta$ is the typical Doppler boost factor
for blazars, probably $\sim 10.$ In the high IGMF
case, this decrease in the contribution from individual blazars
can be partially or completely compensated by the $\delta^2$
times larger number of radio-loud AGN (blazars pointing away from us) 
that are now visible in \g-rays via their cascade radiation. The cascade
backgrounds shown in Fig. 4 apply to the cases when the IGMF is 
very low (the cascade is essentially rectilinear) 
or when the boost in the apparent
blazar luminosity is $\sim \delta^2.$ If one of these cases holds
and the assumptions of Stecker \& Salamon (1996) are correct, then
typical blazar spectra must break strongly at $\sim$ 100
GeV or blazars do not explain the GeV background.
(Note that the precise value of the  break energy depends strongly  
on the IR/O DEBRA, the blazar luminosity function, and the distribution
of blazar \g-ray spectral indices, all of which are 
poorly known.) Ignoring blazars and beaming effects, we can play a 
similar game with quasars as a whole. Assume all quasars isotropically
emit \g-rays with a $dN/dE \propto E^{-2}$ spectrum
up to some maximum energy,
and that, as speculated, their direct emission 
explains the 100MeV-GeV background. If quasar \g-ray emission scales
with optical emission, then using the Boyle {\it et al.} (1991)
 quasar luminosity function, we find 
that for every decade  quasar emission extends above
$\sim 300$ GeV (e.g., \cite{MastiP} predict spectra extending to 
$\sim 10^{16}$ eV for 
radio-quiet AGN), cascading overproduces the background by a 
factor of $\sim 0.25$;
cascading should {\it not} be ignored
in a GeV background calculation involving blazars or quasars.

\section{Conclusion and Discussion}
Since the cascade energy flux falls in the 
same range as the EGRET observations (Fig. 3), we can make a quick
estimate of the maximum average VHE emissivity allowed in the 
Universe by equating
the cascade background energy flux accumulated over a cosmological
distance scale (${1 \over 4\pi} \dot Q_{\rm e-m}d$ where $d \sim
1$ Gpc) with the observed EGRET energy flux above 100 MeV, 
$\sim 8\times 10^3\ {\rm eV}~{\rm cm}^{-2}{\rm sr}^{-1}{\rm s}^{-1}:$
$\dot Q_{\rm em}^{max} \approx 5\times 10^{-23}\ 
{\rm eV}\,{\rm cm}^{-3}{\rm s}^{-1}.$  For source populations
with no or moderate cosmological evolution, the more exact calculation
of Fig. 3 gives an upper limit $\sim 1-3\times 10^{-23} 
\ {\rm eV}\,{\rm cm}^{-3}{\rm s}^{-1}$ for $H_0
=75$ ($\dot Q_{\rm em}^{\rm max} \propto H_0^{-1}$). This is  not a large
number. In a Hubble volume $V_H\sim {4\pi \over 3}(c/H_0)^3,$ this
represents a total VHE source luminosity $\sim 1-3\times10^{50}
{\rm erg}\ {\rm s}^{-1}.$ This can be compared, say, 
to the bolometric luminosity of a single powerful quasar
$\sim 10^{48} {\rm erg}\ {\rm s}^{-1},$  which as discussed,
implies that AGN do not emit much of their luminosity at
VHE energies.
%On intergalaxy scales and
%volumes $\sim 1 {\rm Mpc}^3,$ this represents a total luminosity
%$\sim 1-2\times 10^{39}\ {\rm ergs}s^{-1},$ while the total 
%cosmic ray luminosity of our galaxy is 
%$\sim 10^{41}$ erg/s.
As we have shown, this limit is rather insensitive to 
details of the IR/O background and applies
to {\it any} cosmological population with significant VHE emission above 
$\sim 1$ TeV, e.g., any galaxy or cluster population with
strong cosmic ray production at some stage
in its history.
It will be interesting to see how much of
the GeV \g-ray background future instruments like GLAST can resolve.
If, as suspected, most of the background is non-cascade blazar
emission, the VHE cascade limits tighten and become even more interesting. 
In particular, depending on the IGMF and the details of blazar beaming,
they could imply that typical blazar AGN 
show an {\it intrinsic} 
spectral break (not due to DEBRA absorption) at $\simgreat 100$ 
GeV -- which would be an important constraint for blazar models. 
At the same time, though, a few AGN 
(e.g., Mkn 421) {\it are} effective VHE emitters, and there is no shortage 
of ideas for producing VHE 
emission by other means (e.g., the decaying topological defects discussed
here). Any residual background surviving a fluctuation/point source
analysis could well be VHE cascade emission. Detection
of a cascade background, especially in the cutoff region 
$\simgreat 10$ GeV, provides combined information on the evolution
of the underlying VHE source population  
and the IR/O background (i.e., galaxies). For a low IR/O background,
the cascade background is quite hard. A GLAST detector with sufficient
sensitivity to the diffuse \g-ray background up to 
100 GeV could set even tighter limits on VHE source 
populations (or more easily detect a cascade background).

\section{Acknowledgments} 
\noindent PSC would like to thank the Max-Planck-Institut
f\"ur Kernphysik for its generous hospitality.

\newpage

\centerline{\epsfxsize 6in \epsfbox{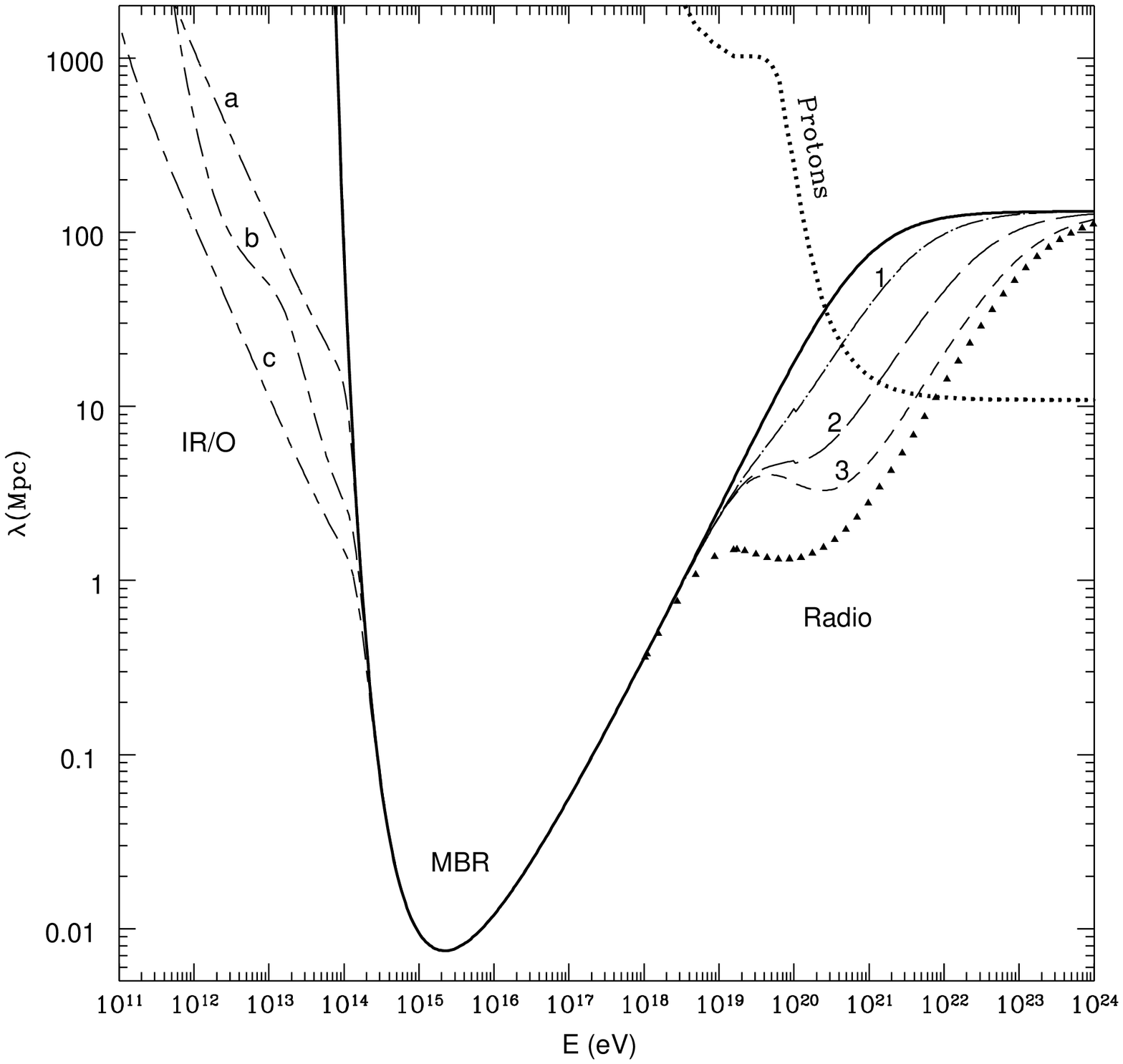}}
\figcaption[fig1.ps]{
The current ($z=0$) pair-production mean free path, $\lambda,$ for VHE photons
of energy, $E.$ 
Below $10^{14}$ eV, VHE photons 
interact primarily with
IR/O photons; above $10^{19}$ eV they interact with radio
photons, and in between, with MBR photons. Curves (a), (b), (c)
 respectively show $\lambda$ for the IR/O backgrounds of
curves (i), (iv), (vi)
in Fig. 2. Curves (1),(2),(3) show respectively $\lambda$ for
the extragalactic radio background estimate of Sironi {\it et. al.} (1990)
(see also Simon 1977) with a low frequency cutoff at 5 MHz, 2 MHz, 
and 1 MHz. The triangles give the lower limit on $\lambda$ obtained 
assuming the total observed radio background (e.g.,
Ressel \& Turner 1991) 
is extragalactic. The {\it heavy dotted} line shows the energy-loss mean 
free path for energetic protons.
% produced by a decaying 
%primordial particle (Geddes, Quinn\& Wald 1996; see \S 3).
\label{fig1}}
\newpage
\centerline{\epsfxsize 6in \epsfbox{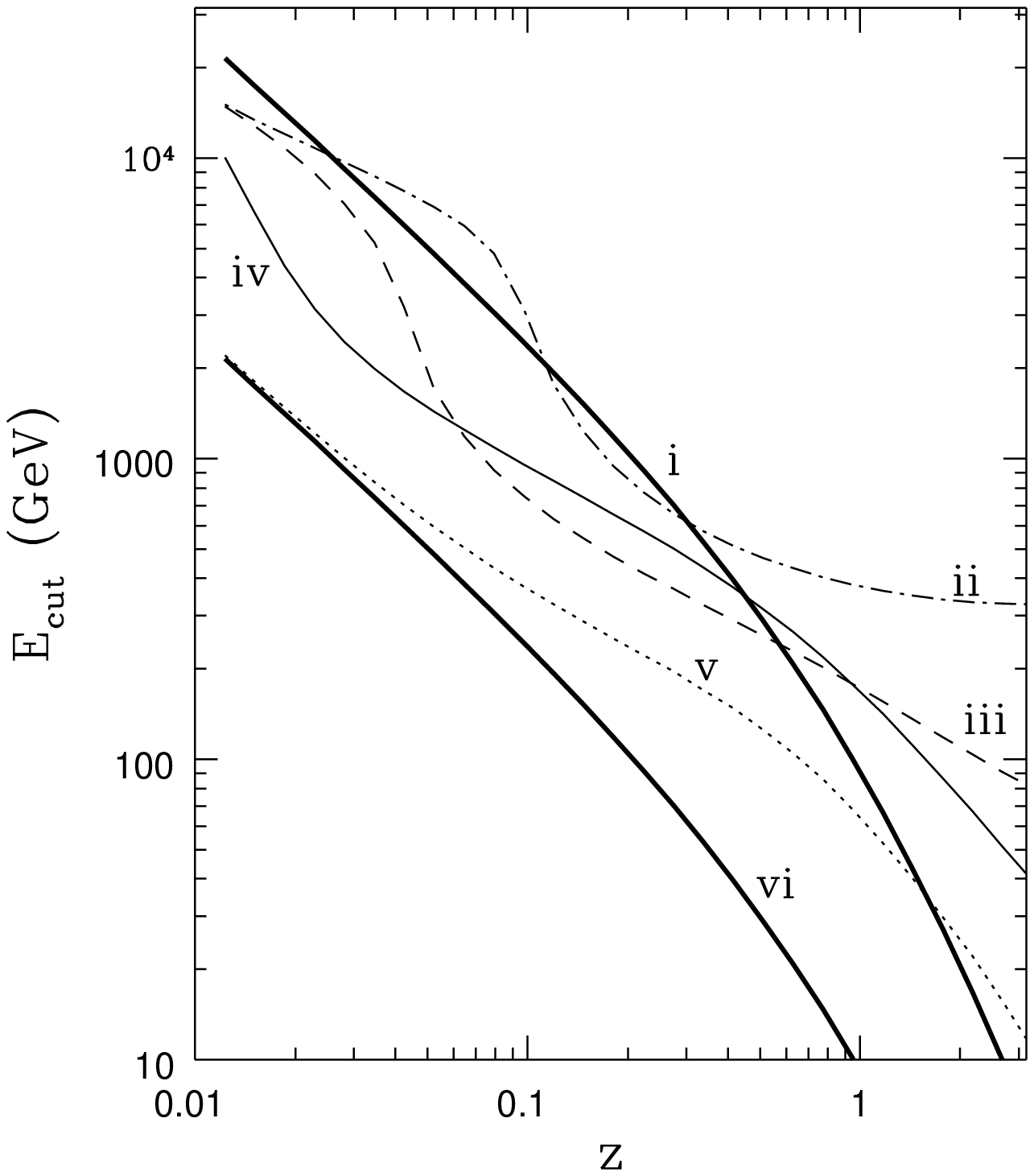}}
\figcaption[fig2.ps]
{ The absorption cutoff energy, $E_{cut},$ as a 
function of source redshift, $z,$ for different IR/O background
models. $E_{cut}$ is defined by the condition $\tau_{\gamma\gamma}
[(1+z)E_{cut},z] = 1,$ where $\tau_{\gamma\gamma}$ is the optical
depth for photon absorption via pair production. See text
for a discussion of the specific DEBRA models shown.
A flat universe with $H_0=75$ was assumed for all calculations.
\label{fig2}}
\newpage
\centerline{\epsfxsize 6in \epsfbox{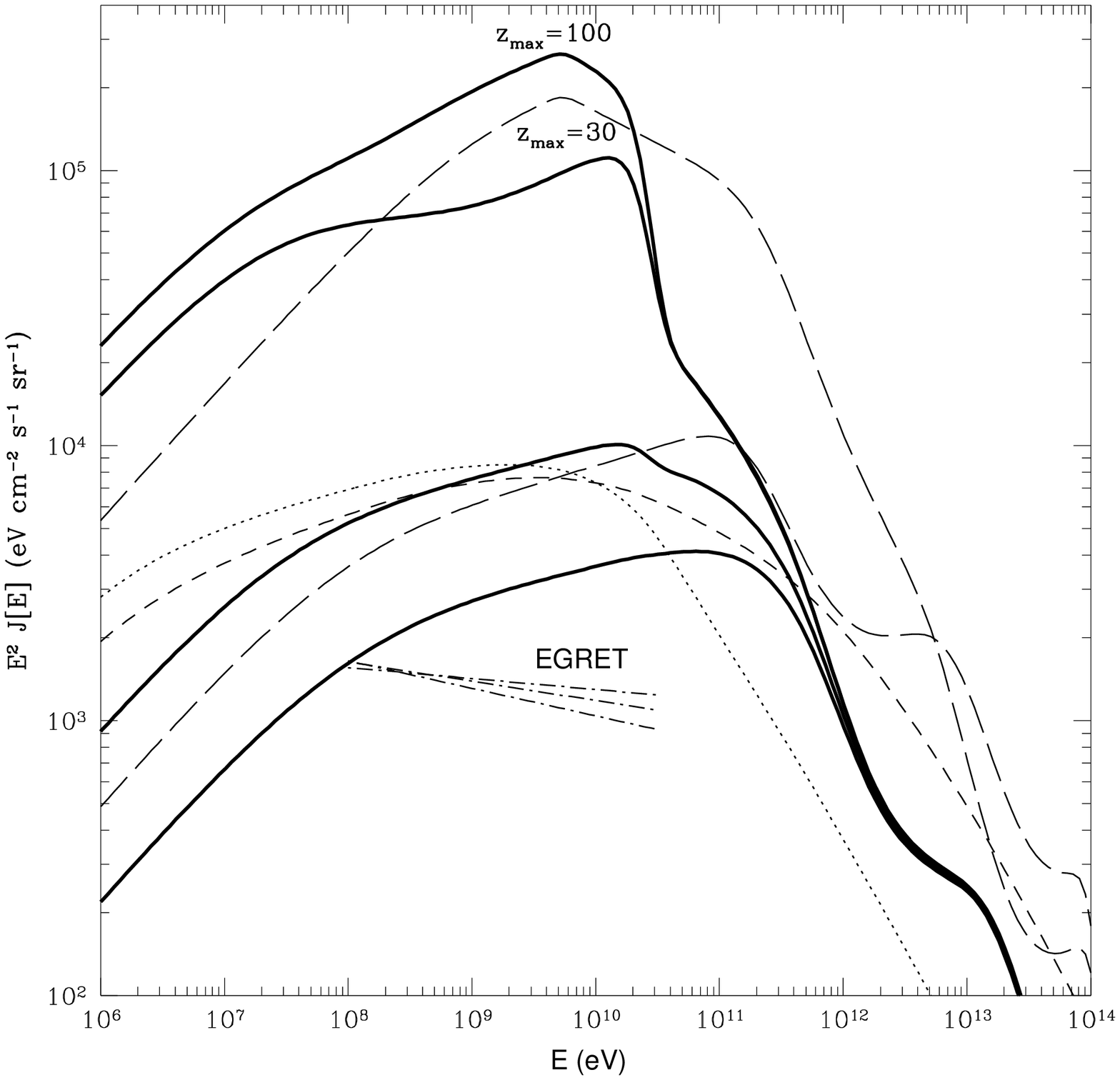}}
\figcaption[fig3.ps]{The VHE cascade background produced for 
different IR/O backgrounds
and topological defect/particle decay rates. From bottom to top, the
{\it heavy, solid} lines show the backgrounds normalized to a
current VHE decay luminosity density of $\dot Q_{e-m}=1\times 10^{-22}\ 
{\rm eV}{\rm cm}^{-3}{\rm s}^{-1}$ that  increases with
redshift as $(1+z)^3,$ $(1+z)^{9/2}$ (the currently favored scenario), and
$(1+z)^{6}.$ The $(1+z)^6$ case does not converge, and we truncate
the integration at two different $z_{max}.$ The IR/O background of 
curve (iv) in Fig. 2 was used. 
The $dotted,$ $dashed$ and $long-dashed$ curves 
respectively show the effects of changing the IR/O background
to that of curves (vi), (i), and (ii) in Fig. 2.
\label{fig3}}
\newpage
\centerline{\epsfxsize 6in \epsfbox{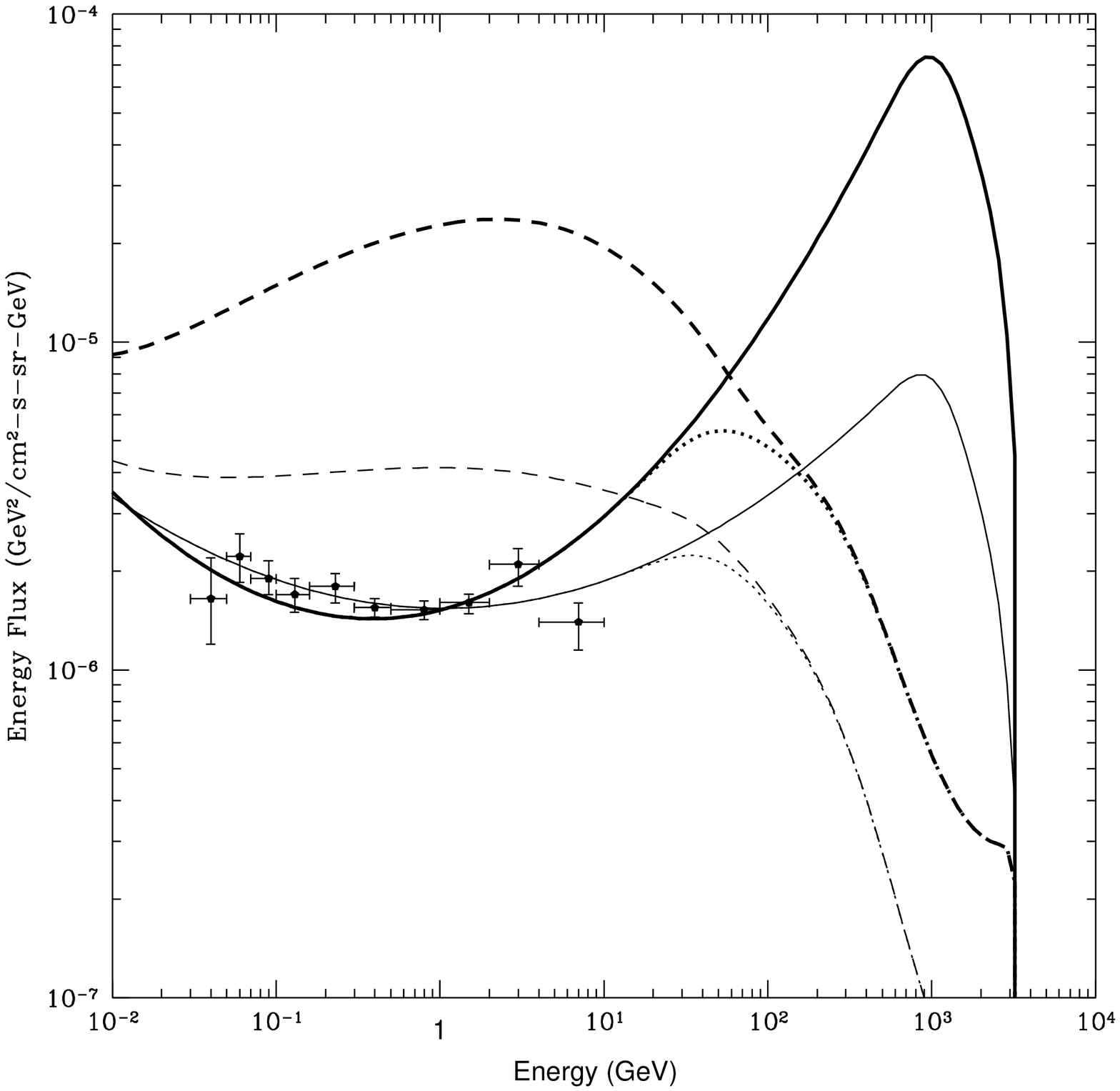}}
\figcaption[fig4.ps]{ The diffuse \g-ray background produced
by blazars. The $solid$ lines show
the same blazar \g-ray background calculation 
as Fig. 3 of Stecker \& Salamon (1996), but assuming
that a blazar spectrum is an unbroken
power law up to 3 TeV and that the blazar spectral index distribution
is a gaussian centered at $\bar \alpha=-2.05$ with 
$\sigma_\alpha=0.25$ ($regular$ weight lines) and 
$\bar \alpha=-2.1$ with $\sigma_\alpha=0.35$ ($heavy$ lines).
The $dotted$ lines show the effects of VHE \g-ray absorption
on the expected background. The $dashed$ lines show the result
when the VHE cascade contribution is included. The IR/O
background of curve (iv) in Fig. 2 was used, but with $H_0=50$
in order to compare with Stecker \& Salamon (1996).
\label{fig4}}
  
\end{document}